\documentclass[twocolumn,prd,nofootinbib,aps,prl,floats,floatfix,amsmath,amssymb,longbibliography,secnumarab
ic]{revtex4-1}

\usepackage[final]{graphicx}
\usepackage{hyperref}
\usepackage{amsmath}
\usepackage{bbm}
\usepackage{amsfonts}
\usepackage{amssymb}
\usepackage{latexsym}
\usepackage{graphicx}
\usepackage[english]{babel}
\usepackage{multirow}
\usepackage{float}
\usepackage{url}
\usepackage{slashed}
\usepackage{xcolor} 
\usepackage[utf8]{inputenc}
\usepackage{verbatim}
\usepackage{stackengine}

\def\sfrac#1#2{{\textstyle{#1\over #2}}}
\newcommand{\be}{\begin{equation}}
\newcommand{\ee}{\end{equation}}
\newcommand{\ba}{\begin{array}}
\newcommand{\ea}{\end{array}}
\newcommand{\bea}{\begin{eqnarray}}
\newcommand{\eea}{\end{eqnarray}}
\newcommand{\sss}{\scriptscriptstyle}

\newcommand{\nn}{\nonumber}
\newcommand{\A}{{\sss A}}

\newcommand{\W}{{\sss W}}
\newcommand{\Z}{{\sss Z}}
\newcommand{\F}{{\sss F}}

\newcommand{\diff}{\mathrm{{d}}}

\begin{document}

\title{
Sterile neutrino production at small mixing in the early universe}

\author{Gonzalo Alonso-\'Alvarez}
\affiliation{McGill University, Department of Physics, 3600 University St.,
Montr\'eal, QC H3A2T8 Canada}
\author{James M.\ Cline}
\affiliation{McGill University, Department of Physics, 3600 University St.,
Montr\'eal, QC H3A2T8 Canada}
\begin{abstract}

Sterile neutrinos can be produced in the early universe via interactions with their active counterparts.
For small active-sterile mixing angles, thermal equilibrium with the standard model plasma is not reached and sterile neutrinos are only produced via flavor oscillations.
We study in detail this regime, taking into account matter potentials and decoherence effects caused by elastic scatterings with the plasma.
We find that resonant oscillations occurring at temperatures $T\lesssim 10$\,GeV lead to a significant enhancement of the sterile neutrino production rate.
Taking this into account, we improve constraints on the active-sterile mixing from Big Bang nucleosynthesis and the cosmic microwave background, excluding mixing angles down to $\theta_s\sim 10^{-10}-10^{-16}$ for sterile neutrino masses in the $10$~MeV to $10$~GeV range.
We observe that if sterile neutrinos predominantly decay into metastable hidden sector particles, this process provides a novel dark matter production mechanism, consistent with the sterile neutrino origin of light neutrino masses via the seesaw mechanism.
\end{abstract}

\maketitle

{\bf Introduction.}
Sterile neutrinos ($\nu_s$) are among the simplest and most natural extensions of the standard model (SM).
The sterile neutrino flavors are inevitably produced in the early Universe through oscillations with active neutrinos ($\nu_a$)~\cite{Barbieri:1989ti,Kainulainen:1990ds,Enqvist:1990ad,Barbieri:1990vx,Enqvist:1990ek,Enqvist:1991qj}.
This was used to derive constraints on the sterile neutrino mass $m_s$ and mixing angle $\theta_s$ from Big Bang nucleosynthesis (BBN).
In the seminal papers, resonant oscillations were shown to occur for sterile neutrino masses smaller than the
active neutrino ones, which at that time were much less stringently constrained than now. 
In this work we are concerned with the opposite case, of relatively heavy $\nu_s$ with $m_s\gtrsim 10\,$MeV.
Sterile neutrinos around the GeV scale are not only well-motivated theoretically~\cite{Asaka:2005an,Asaka:2005pn,Ghiglieri:2017gjz,Ghiglieri:2017csp,Ghiglieri:2018wbs,Klaric:2020lov,Bondarenko:2021cpc}, but also interesting from phenomenological and experimental perspectives~\cite{Bondarenko:2018ptm}.

More recently, BBN has been used to place bounds on heavier and more strongly coupled sterile neutrinos which can be produced via thermal processes in the early universe plasma~\cite{Dolgov:2000jw,Dolgov:2000pj,Dolgov:2003sg,Ruchayskiy:2012si,Boyarsky:2020dzc,Gelmini:2020ekg,Sabti:2020yrt}.
The regime of small mixing $\theta\lesssim10^{-6}$ for which thermalization does not occur has garnered less attention, and in this work we aim to fill that gap.
Aside from BBN, the cosmic microwave background (CMB) temperature fluctuations are also very sensitive to exotic particle species, especially if they can decay electromagnetically~\cite{Slatyer:2016qyl,Poulin:2016anj}.
Ref.~\cite{Poulin:2016anj} studied the impact of a decaying sterile neutrino population on the CMB, however without making any connection with their production mechanism.

An appealing possibility is that a cosmological population of sterile neutrinos makes up the dark matter (DM) of the universe.
Ref.~\cite{Dodelson:1993je} 
proposed that nonresonant oscillations could explain the origin of $\nu_s$ as a warm dark matter candidate, and showed it to be consistent with BBN constraints as long as $m_s\gtrsim 100\,$eV. 
Although this mechanism also works for larger masses \cite{Abazajian:2017tcc}, it is now constrained to $m_s\lesssim 3\,$keV by X-ray searches for the decay $\nu_s\to\gamma+\nu_a$ \cite{Boyarsky:2018tvu}. 
Such light $\nu_s$ are in the warm dark matter regime strongly disfavored by Lyman-$\alpha$ observations \cite{Yeche:2017upn}. 
Thus, the simplest version of $\nu_s$ as a dark matter candidate seemed to be ruled out.

A possible loophole is to postulate a large cosmic lepton
asymmetry \cite{Shi:1998km,Laine:2008pg}, allowing the oscillations to be resonant and somewhat relaxing the X-ray constraints on $\theta_s$.  But even in this case the allowed parameter space window is small \cite{Perez:2016tcq}, considering
the constraints on the lepton asymmetry from BBN and on $m_s$ from structure formation \cite{Abazajian:2001nj}. 
Hence, various other alternatives going beyond the minimal scenario have been considered for the production of $\nu_s$ as dark matter~\cite{Kusenko:2006rh,Petraki:2007gq,Adhikari:2016bei,Kusenko:2010ik,Shuve:2014doa,Alonso-Alvarez:2021pgy,Chao:2021grp}. 
Ref.~\cite{Datta:2021elq} recently argued that sterile neutrino dark matter with  $m_s\lesssim 1\,$MeV can be produced through freeze-in, with no need for oscillations or additional new physics.

In this paper, we explore a regime for resonantly producing GeV-scale $\nu_s$ that neither requires additional new physics beyond the standard model, nor relies on the existence of any background lepton asymmetry.\footnote{This resonance was first noticed in Ref.~\cite{Ghiglieri:2016xye}.}  
Instead, it takes advantage of the fact that the neutrino matter potential changes sign at temperatures below the electroweak phase transition.
Due to the matter potential, the active neutrinos have a self-energy that can exceed $m_s$ above $T\sim 10\,$GeV. 
A crossing of levels occurs as $T$ decreases, leading to resonant enhancement of the oscillations.  
The precise temperature at which the resonance occurs depends on the energy of the active neutrino.
We demonstrate that this produces a cosmologically relevant $\nu_s$ population for mixing angles as small as $\theta_s \sim 10^{-10}-10^{-16}$,
for sterile neutrinos in the mass range $m_s \sim 0.01-10$\,GeV.

Even for such small mixing angles, the decays of $\nu_s$ are too fast for this population to constitute the dark matter of the universe.
However, the hadronic and electromagnetic decay products can distort the temperature fluctuations of the CMB and alter the light-element yield predictions of BBN.
We find that the resonant
production mechanism leads to new stringent upper bounds on $\theta_s$.
These might be circumvented in a more complicated dark sector, in which $\nu_s$ decays predominantly to some other hidden particle.  If the decay product is massive and metastable on cosmological time scales, it could constitute the
dark matter of the universe, as we will show.

\bigskip
{\bf Resonant $\nu_s$ oscillations.}  For simplicity, we consider oscillations of $\nu_s$ with a single flavor of active neutrinos, $\nu_a$, where $a = e$, $\mu$, or $\tau$.  For relativistic neutrinos the oscillations can be determined by solving the Schr\"odinger equation for neutrinos of momentum $p$, with the
$2\times 2$ Hamiltonian
\be
    H = p+ \frac{1}{2p}\left({m_{aa}^2\atop m_{as}^2}\, {m_{as}^2\atop m_{ss}^2}\right)
    +\left({V_a\atop 0}\,{0\atop 0}\right)
    \label{Heq}
\ee
in the flavor basis $(\nu_a,\,\nu_s)$. 
Here, $m^2_{ij}$ are elements of the neutrino mass-squared matrix, which can be diagonalized through a rotation of angle $\theta_s \simeq m_{as}^2/m_{ss}^2$, known as the vacuum mixing angle.
Since $\theta_s$ will turn out to be small in the relevant regions of parameter space, $m_{ss}^2 = m_s^2$ to a good approximation.

\begin{figure}[!t]
\begin{center}
\includegraphics[width=0.99\linewidth]{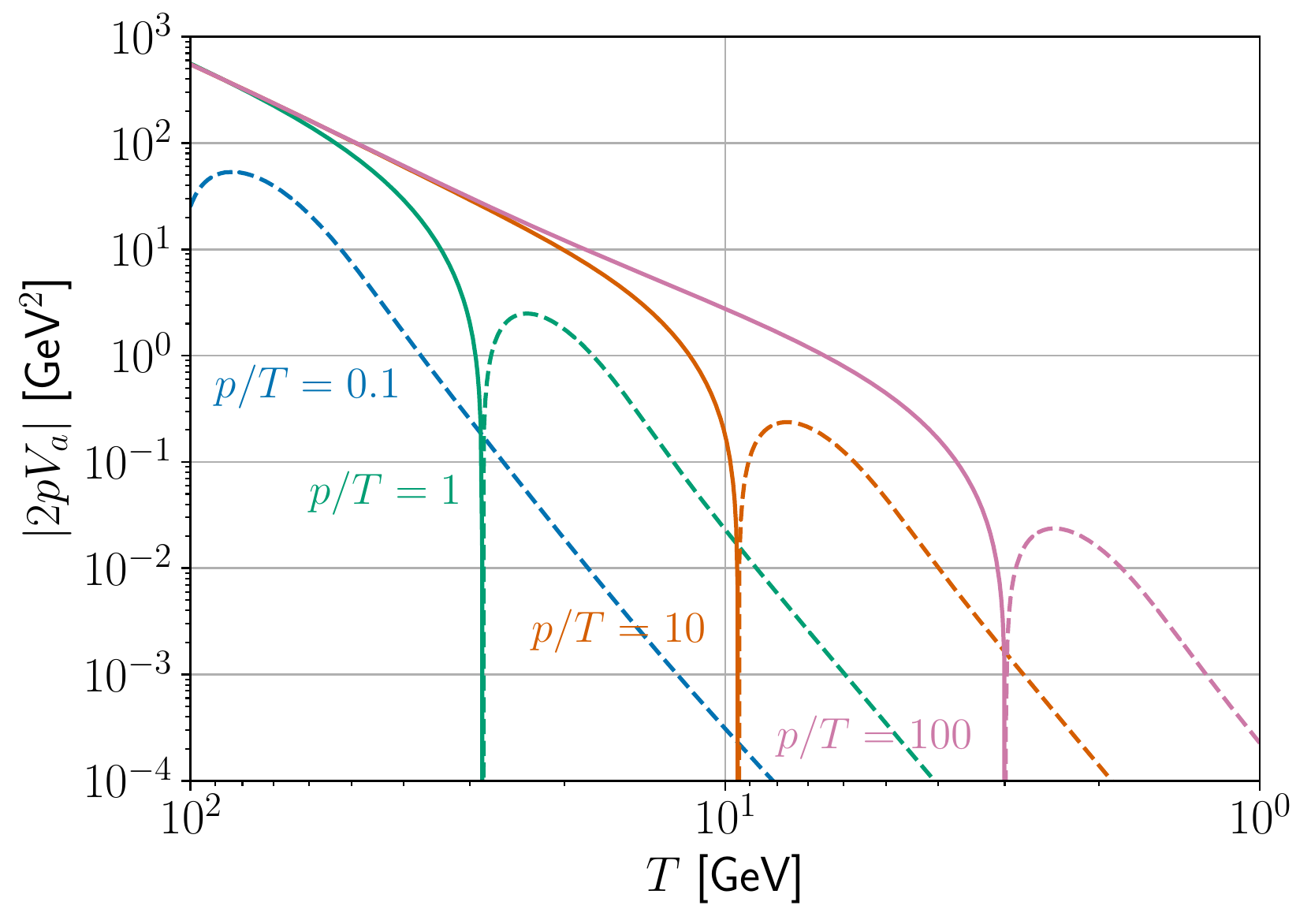}
\vskip-0.5cm
\caption{Absolute value of $V_a$,  the matter potential for active neutrinos, times twice their momenta $p$, as a function of the temperature of the Universe.
Solid (dashed) lines indicate where $V_a$ is positive (negative).
The resonance occurs  when $2p V_a$ crosses $m_s^2$ which occurs close where $V_a$ changes sign.}
\label{fig:Va}
\end{center} 
\end{figure}

The matter potential $V_a$ arises from well-known thermal corrections \cite{Weldon:1982bn,Notzold:1987ik, Quimbay:1995jn}, which in the high- and low-temperature limits are approximately given by
\be
    V_a \simeq \left\{\begin{array}{ll}\!\!- 14\pi/ (45\alpha)\,(2\mathrm{e}^{-\tfrac{m_\ell}{T}}+c_\W^2)s_\W^2\,G_\F^2 T^4 p,  & T\ll T_{\rm EW}\\
 \!\! \phantom{|^|} (3 g_2^2+g_1^2)\,T^2/(32\, p), & T \gg T_{\rm EW}
    \end{array}\right.
    \label{eq:vweq}
\ee
where $g_i$ are the SM gauge couplings, $s_\W$ and $c_\W$ are the sine and cosine of the Weinberg angle, $\alpha \simeq 1/137$ is the fine structure constant, and $m_{\ell}$ is the mass of the charged lepton of corresponding flavor.
The crucial sign change in $V_a$ mentioned above is evident in Eq.~\eqref{eq:vweq}, which we show only for illustration.  For quantitative purposes, the
exact dependence $V_a(p,T)$ that smoothly connects these approximations is needed.
Furthermore, for neutrinos undergoing active-sterile oscillations, the self-energy has to be evaluated at the on-shell point for the heavy sterile states.
The detailed calculations are described in the Appendix and the resulting matter potential is depicted there as a function of $T$ (Fig.~\ref{fig:Va}).

The Hamiltonian~\eqref{Heq} including the matter potential is diagonalized with a mixing angle in matter $\theta_m$ given approximately by 
\be
\sin^2(2\theta_m) \simeq \frac{4\theta_s^2}{4\theta_s^2 + \left(2p V_a/m_s^2 - 1\right)^2}\,,
\label{mixang}
\ee
in the limit $m_{aa}^2, m_{as}^2\ll m_s^2$. 
This displays a resonance at $2pV_a = m_s^2$, which for $m_s\ll T$ is close to where $V_a$ crosses zero.    
For given values of $T$ and $m_s$, it occurs when the dimensionless momentum $q=p/T$ satisfies $q=q_r$, the latter being defined by 
\be
2q_r V_a(q_r,T) = {m_s^2/T}\,.
\ee
Although the resonance in (\ref{mixang}) appears to be narrow, it becomes broad when the elastic scatterings of active neutrinos in the plasma are taken into account.

The time-dependent active-to-sterile oscillation probability enabled by the nonzero mixing is given by
\begin{equation}\label{eq:instantaneous_oscillation_probability}
P_{\nu_a\to\nu_s} = \sin^2(2\theta_m)\sin^2(\Delta E\,t/2)\,,
\end{equation}
where $\Delta E \simeq V_a - m_s^2/(2p)$ is the energy splitting between the two states (neglecting the small active neutrino mass). 
Elastic scatterings of active neutrinos with the plasma act to decohere the $\nu_a\rightarrow\nu_s$ oscillations.
This can be taken into account by averaging the instantaneous oscillation probability~\eqref{eq:instantaneous_oscillation_probability} over the interaction time~\cite{Barbieri:1989ti,Abazajian:2001nj,Bringmann:2018sbs}, to give the mean oscillation probability
$\bar P_{\nu_a\to\nu_s}$.  The rate of $\nu_a\to\nu_s$ conversions, taking into account decoherence, is then
\bea
\Gamma_{a\to s} &=& 
    \Gamma_a \bar P_{\nu_a\to\nu_s} = \Gamma_a^2\int \mathop{\diff t}\, P_{\nu_a\to\nu_s} \mathrm{e}^{-\Gamma_a t}\nn\\
    &=& \frac{\Gamma_a}{2}\frac{\theta_s^2 m_s^4}{
  \left(p V_a - m_s^2/2\right)^2+(p\Gamma_a/2)^2}\,,
  \label{eq:averaged_oscillation_rate}
\eea
where we have used the fact that $\theta_s \ll p \Gamma_a/m_s^2$ for the range of parameters of interest.
From this we see that the width of the resonance is in fact determined by $p\Gamma_a/m_s^2$ rather than $\theta_s$.  

The elastic scattering rate is often approximated by~\cite{Notzold:1987ik}
\be
    \Gamma_a \simeq  \frac{91\pi}{ 216}G_\F^2T^4 p \equiv C_a^2 T^4 p
\ee
at temperatures $m_{\ell} < T \ll T_{\rm EW}$, but in general it has more a complicated dependence on $p$ and $T$.  
For our numerical evaluations, we use the tabulated values provided in Refs.\ \cite{Asaka:2006nq,Ghiglieri:2016xye}.

\bigskip
{\bf Abundance of $\nu_s$.}
The relative abundance between sterile and active neutrinos of momentum $p$ at a given temperature, denoted by $R(p,T)$, satisfies the Boltzmann equation 
\begin{equation}
\frac{\diff R}{\diff t} = \Gamma_{a\to s}(1-R)\,,
\end{equation}
whose solution when $R\ll 1$ is
\bea
    R(p,T) &\simeq& \int_T^\infty \mathop{\diff T'} \frac{\Gamma_{a\to s}}{ H(T')T'}\, {\frac{g_{*s}(T_*)}{g_{*s}(T')}}\,.
\label{eq:Tint}
\eea
Here, $T_*$ is defined as $T_*=\max{(T,1\,\mathrm{MeV})}$ so that the ratio of relativistic degrees of freedom takes into account the dilution of the sterile neutrinos due to entropy dumps into the thermal bath after their production, but before active neutrino decoupling.
The total abundance of sterile neutrinos
relative to active ones is obtained by the phase space integration
\be
Y_{s:a} = \frac{\int \mathop{\diff^3 p} R(p,T)\, n_f(p/T)}{\int \mathop{\diff^3 p} n_f(p/T)}\,,
\label{phsp}
\ee
where $n_f(p/T)$ is the Fermi-Dirac distribution function.  For a narrow resonance, it would be possible to do one of the integrals analytically
using the narrow width approximation, but here we must carry them out numerically.
Defining the integral
\begin{align}
    &I(m_s, T) = \nn\\ 
    &\int_T^\infty \mathop{\diff T'} \int \mathop{\diff q} 
    {[g_{*s}(T_*)/g_{*s}(T')]\,\Gamma_a q^2\,T'^{-3} \over \sqrt{g_*}\, (\mathrm{e}^q+1)\left[\left(q T' V_a -\sfrac12 m_s^2\right)^2 + \gamma^2\right]},
    \label{Ieq}
\end{align}
where $\gamma = q T'\, \Gamma_a/2$,\footnote{In principle there can be further contributions to the damping rate entering $\gamma$, including the decays of $\nu_s$ and the differential spreading of the $\nu_s$ and $\nu_a$ wave packets that leads to decoherence.  Numerically we find that these are negligible for the parameters of interest.} we can express the relative abundance of sterile versus active neutrinos as
\be
    Y_{s:a}(T) = {2\, m_s^4\,\theta_s^2\, M_p\over 3\zeta(3)\cdot 1.66}\, {I(m_s,T)}.
    \label{Yseq}
\ee
For sufficiently small $T\ll 1$\,MeV, $I$ and $Y_{s:a}$ become independent of $T$.  We denote $I(m_s)\equiv I(m_s,T\ll 1\,\mathrm{MeV})$ and similarly for $Y_{s:a}$ at these low temperatures.

\begin{figure}[t]
\begin{center}
 \includegraphics[width=0.925\linewidth]{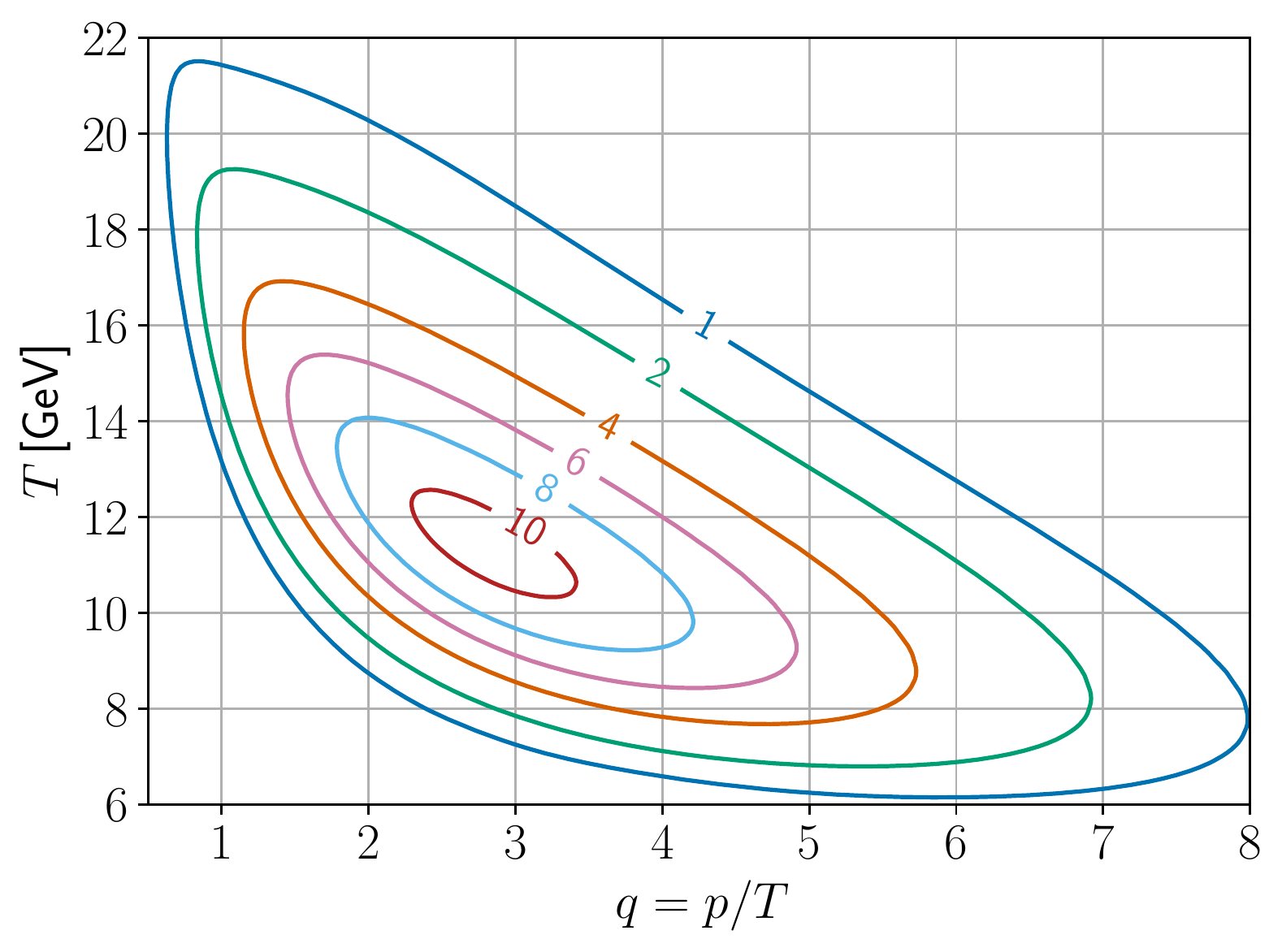}
\vskip-0.5cm
 \caption{Contours of the integrand in Eq.\ (\ref{Ieq}), showing the localization of the resonance in momentum and temperature, for a sterile neutrino mass $m_s=1\,$GeV.  Contours are labeled in units of $10^{-8}$\,GeV$^{-6}$.}
 \label{fig:cont}
\end{center} 
\end{figure}

The abundance relative to the entropy $s$ is
\begin{equation}
Y_s = \frac{n_{\nu}}{s}\,Y_{s:a} \simeq \frac{3}{77}\,Y_{s:a}\,.
\end{equation}

\begin{figure}[t]
\begin{center}
\includegraphics[width=\linewidth]{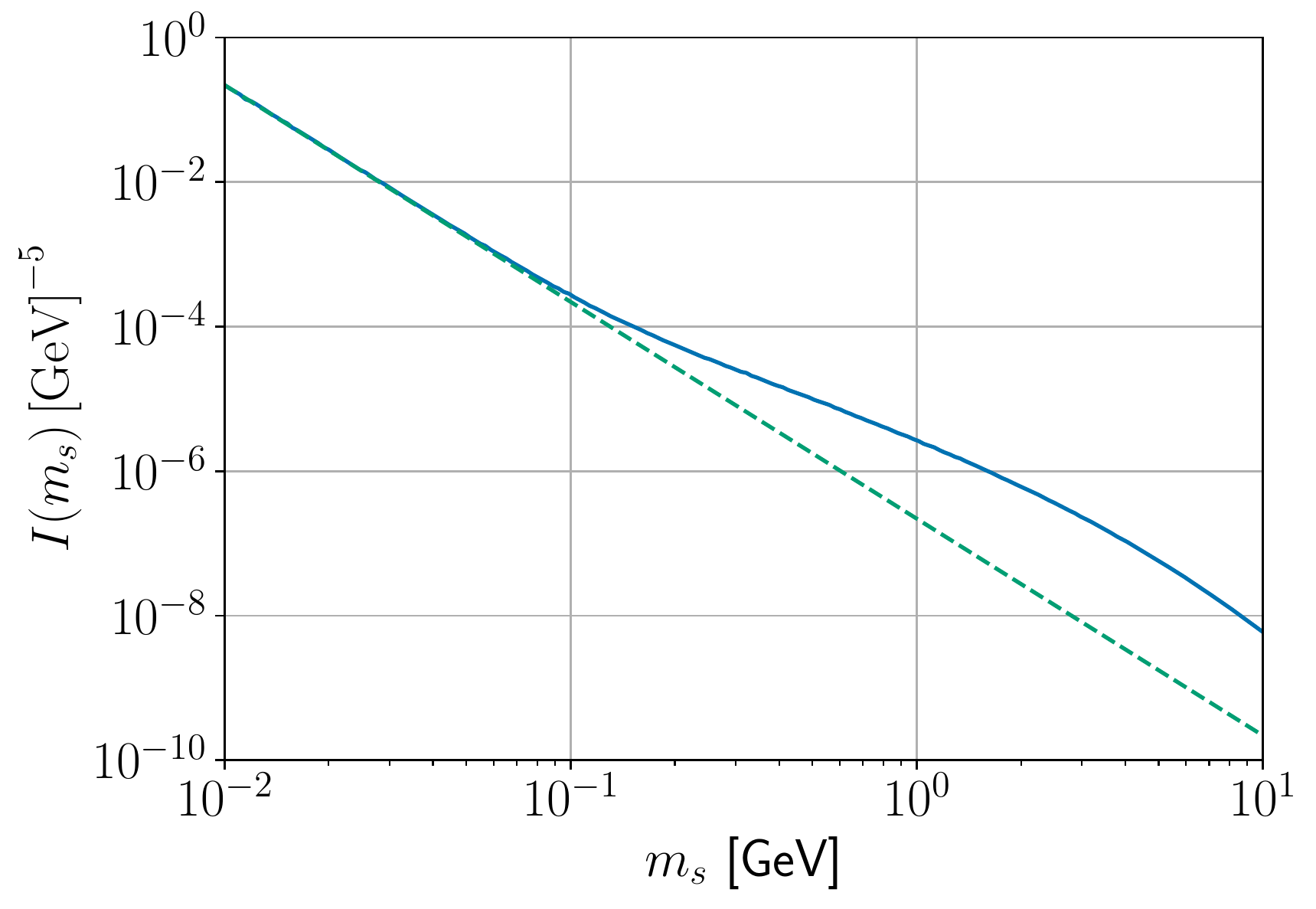}
\vskip-0.5cm
\caption{The integral $I$ that determines the $\nu_s$ abundance, shown as a function of the sterile neutrino mass. The dashed line is the extrapolation of the low-$m_s$ part of the curve, where the resonance plays no role.}
 \label{fig:igr}
\end{center} 
\end{figure}

Contour levels of the integrand
in Eq.\ (\ref{Ieq}) are shown in Fig.\ \ref{fig:cont}, for an exemplary sterile neutrino mass of $m_s=1\,$GeV. 
The resonance peak is pronounced, despite being broad.
Carrying out the integral over a range of $m_s$ yields
the result for $I(m_s)$ shown in Fig.\ \ref{fig:igr}.
The dashed line extrapolates values of $I(m_s)$ at low
masses where the resonance is not effective. The difference shows that the resonance can enhance the $\nu_s$ abundance by over an order of magnitude compared to production by nonresonant oscillations.
Given that the resonant production occurs at temperatures much larger than any charged lepton mass, this result is effectively independent of the flavor of the active neutrino.

A more sophisticated way to derive the $\nu_s$ abundance from oscillations is the density matrix formalism (see, for example,~\cite{Asaka:2006rw}).
Our more phenomenological approach has the appeal of being simple and physically intuitive.
To check the accuracy of our method, we have compared its predictions for the $\nu_s$ abundance in the nonresonant regime with those of the density matrix formalism~\cite{Asaka:2006nq}, finding agreement within the error bars presented there.

\bigskip
{\bf CMB and BBN constraints.}  The sterile neutrinos produced
in the way described above cannot be dark matter, since the values of $\theta_s$ required to reproduce the observed DM abundance lead to $\nu_s$ decays on time scales much shorter than the age of the universe:
\be
    \tau_s  \sim (\theta_s G_F)^{-2} m_s^{-5}
    \sim 10^{-12}\,\theta_s^{-2}\left({\rm GeV}\over m_s\right)^5\,{\rm s}.
    \label{Gweak}
\ee   
For GeV-scale $\nu_s$ a more precise estimation can be obtained rescaling the decay rate of the tau lepton, which is what we do to obtain the numerical value in~\eqref{Gweak}.

The mixing induces weak decays into $\gamma + \nu_a$ at one loop, and into neutrinos, leptons, or quarks at tree level.
For $m_s\gtrsim 1\,$GeV, the branching ratio into hadrons is 
$\sim 0.7$ \cite{Bondarenko:2018ptm}.
The injection of such decay products at the time of CMB formation leads to distortions of the temperature fluctuations, whose nonobservation constrains the product $m_s Y_s\propto \rho_s$
(times the efficiency of electromagnetic energy injection) as a function of the lifetime $\tau_s$ \cite{Slatyer:2016qyl,Poulin:2016anj}.
Similarly, electromagnetic or hadronic decays before and during BBN can affect the successful predictions for the light element abundances~\cite{Poulin:2015opa,Kawasaki:2017bqm}.

\begin{figure}[t]
\begin{center}
 \includegraphics[width=\columnwidth]{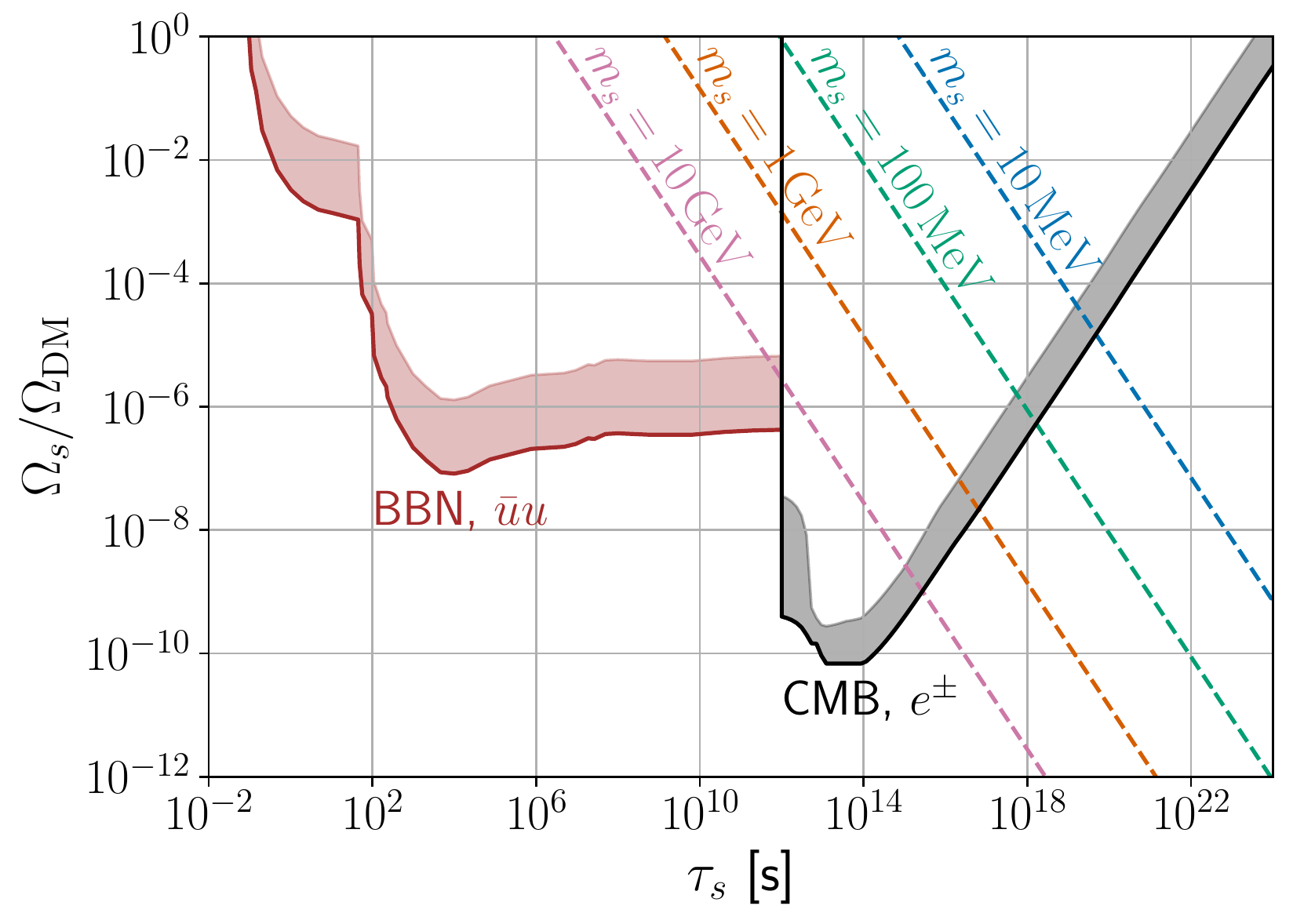}
 \vskip-0.5cm
 \caption{Abundance relative to the observed DM one versus lifetime of the sterile neutrino, showing predictions from production through oscillations (diagonal dashed lines) and limits from BBN (hadronic injection) and CMB ($e^\pm$ injection). The width of the bands in the constraints shows how the limits vary within the range of $m_s$ considered.}
 \label{fig:bbn}
\end{center} 
\end{figure}

For a given value of the sterile neutrino mass, the relation between $m_s Y_s$ and $\tau_s$ is parametrized by $\theta_s$, giving diagonal contours in Fig.~\ref{fig:bbn}, which displays $\Omega_s / \Omega_{\rm DM} \propto m_s Y_s$ on the vertical axis.
Fig.~\ref{fig:bbn} exhibits the upper limits from energy injection at CMB formation from Ref.\ \cite{Poulin:2016anj} and at BBN from Ref.\ \cite{Kawasaki:2017bqm}.
The efficiencies of the respective electromagnetic and hadronic injections are implemented in our limits, using the predictions of Ref.\ \cite{Bondarenko:2018ptm} for the relevant branching ratios of the sterile neutrino as a function of its mass.

The CMB limit arises through the distortion of the temperature fluctuations by injection of $e^\pm$ from $\nu_s$ decays. 
This limit is somewhat stronger than that coming from $\gamma$ injection.
The grey-shaded region in Fig.~\ref{fig:bbn} encompasses the ensuing limits as the kinetic energy of the injected particles ranges from $10~$keV to $1~$TeV, and is representative of the variance of the limit for the values of $m_s$ that we consider.
Note that the CMB temperature fluctuations do not constrain sterile neutrinos with lifetimes shorter than $\sim 10^{12}~$s.
Although the limits from CMB spectral distortions~\cite{Ellis:1990nb, Hu:1993gc,Chluba:2013wsa,Chluba:2013pya,Poulin:2016anj} can constrain shorter lifetimes, they are significantly weaker than the BBN constraints that we describe next.

The BBN constraints shown in Fig.~\ref{fig:bbn} result from the injection of hadrons from $\nu_s$ decays, specifically $u\bar{u}$ quark pairs, as considered in Ref.\ \cite{Kawasaki:2017bqm}.
They are typically stronger than those based on electromagnetic energy injections, but they are only applicable for sufficiently heavy sterile neutrinos, such that hadronic decays are kinematically allowed.
The red-shaded region in Fig.~\ref{fig:bbn} takes into account the variation of the hadronic branching fractions as $m_s$ goes from $1$ to $10~$GeV.

For small mixing and in the range of masses under consideration, BBN gives weaker limits than the CMB.
Nevertheless, BBN would become more constraining for large $m_s$, for which $\tau_s$ drops below $10^{12}$~s. 
This however only occurs for $m_s\gtrsim 100\,$GeV, which is outside of the range of validity of our approximations, as such massive $\nu_s$ would not be relativistic at temperatures $T\lesssim 10$~GeV.
The CMB limits do  not apply for intermediate values of $\theta_s$ for which the lifetime shrinks below $\tau_s < 10^{12}$\,s.
As shown in Fig.~\ref{fig:bbn}, the BBN bounds take precedence for those intermediate mixings, leaving no unconstrained region for $m_s\leq 10$~GeV.

\begin{figure}[t]
\begin{center}
\vskip-0.30cm
\centerline{$\!\!\!\!\!$\includegraphics[width=1.05\linewidth]{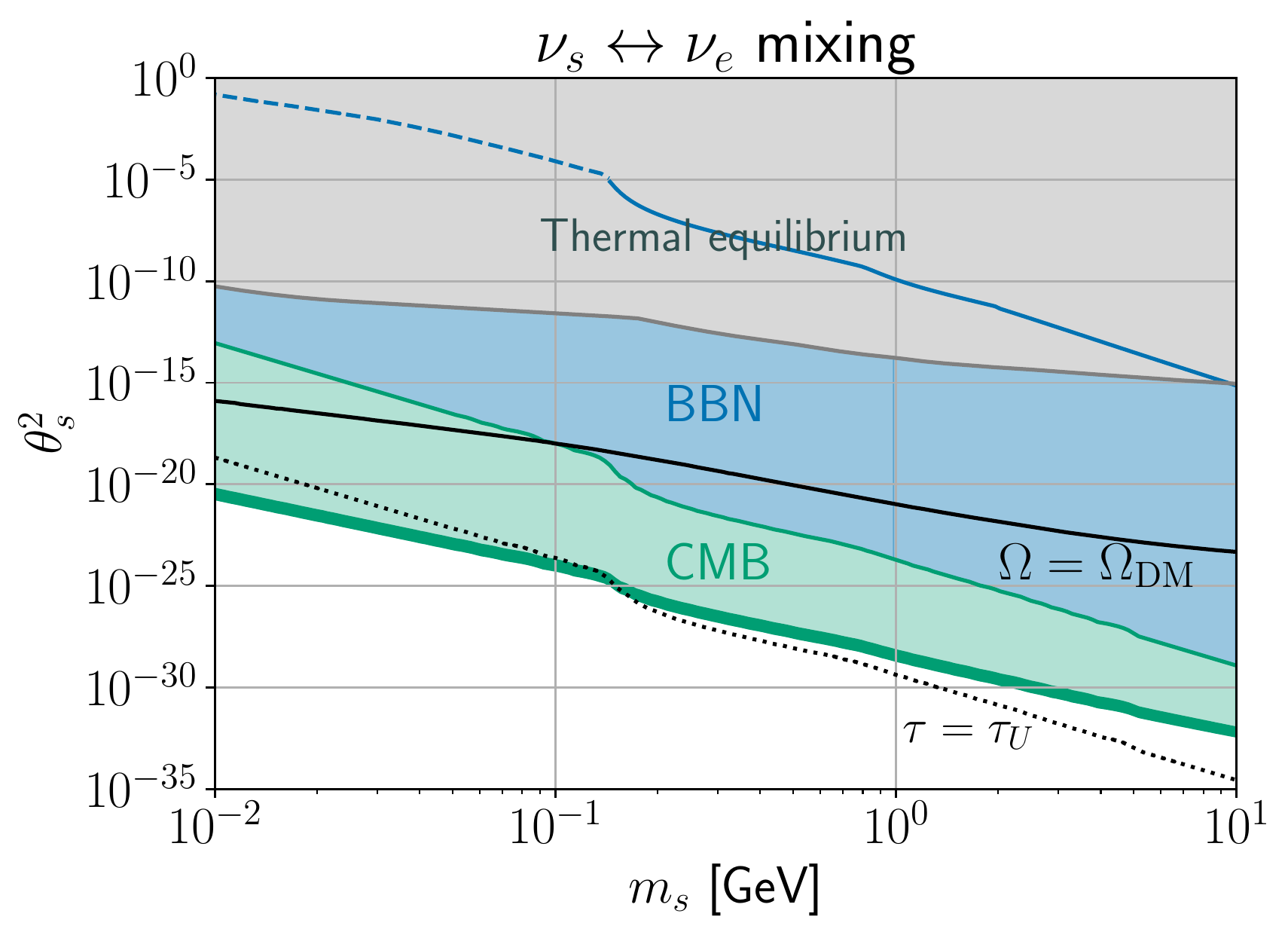}}
\vskip-0.5cm
\caption{BBN (blue) and CMB (green) limits on $\nu_s$ mixing with $\nu_e$ as a function of $m_s$.
The solid black line indicates where the abundance of the produced $\nu_s$ equals the observed dark matter one ($\nu_s$ are however not stable there, the dotted curve indicates a lifetime matching the age of the universe $\tau_U$).
In the gray region the $\nu_s$ are thermally produced and our analysis is not applicable. 
Solid and dotted blue lines denote the upper reach of BBN constraint (see text for details).}
 \label{fig:bbn_e}
\end{center} 
\end{figure}

By determining where the dashed curves intersect with the respective limit in Fig.~\ref{fig:bbn}, one can translate the CMB and BBN bounds to the $m_s$-$\theta_s$ parameter space,  which is shown in Fig.~\ref{fig:bbn_e}.
As expected, CMB temperature fluctuations constrain the smaller values of active-sterile neutrino mixing, while BBN constraints take over for intermediate values of $\theta_s$.
For reference, the solid black line shows the combinations of $m_s$ and $\theta_s$ for which a density equal to the observed dark matter one is generated via resonant oscillations. 
However, these mixings render the $\nu_s$ lifetime shorter than the age of the Universe (the dotted line in Fig.~\ref{fig:bbn_e} shows where the two are equal). 
This means that $\nu_s$ cannot possibly be a dark matter candidate in this simplest setup.

\begin{figure*}[t]
\begin{center}
 \includegraphics[width=0.45\linewidth]{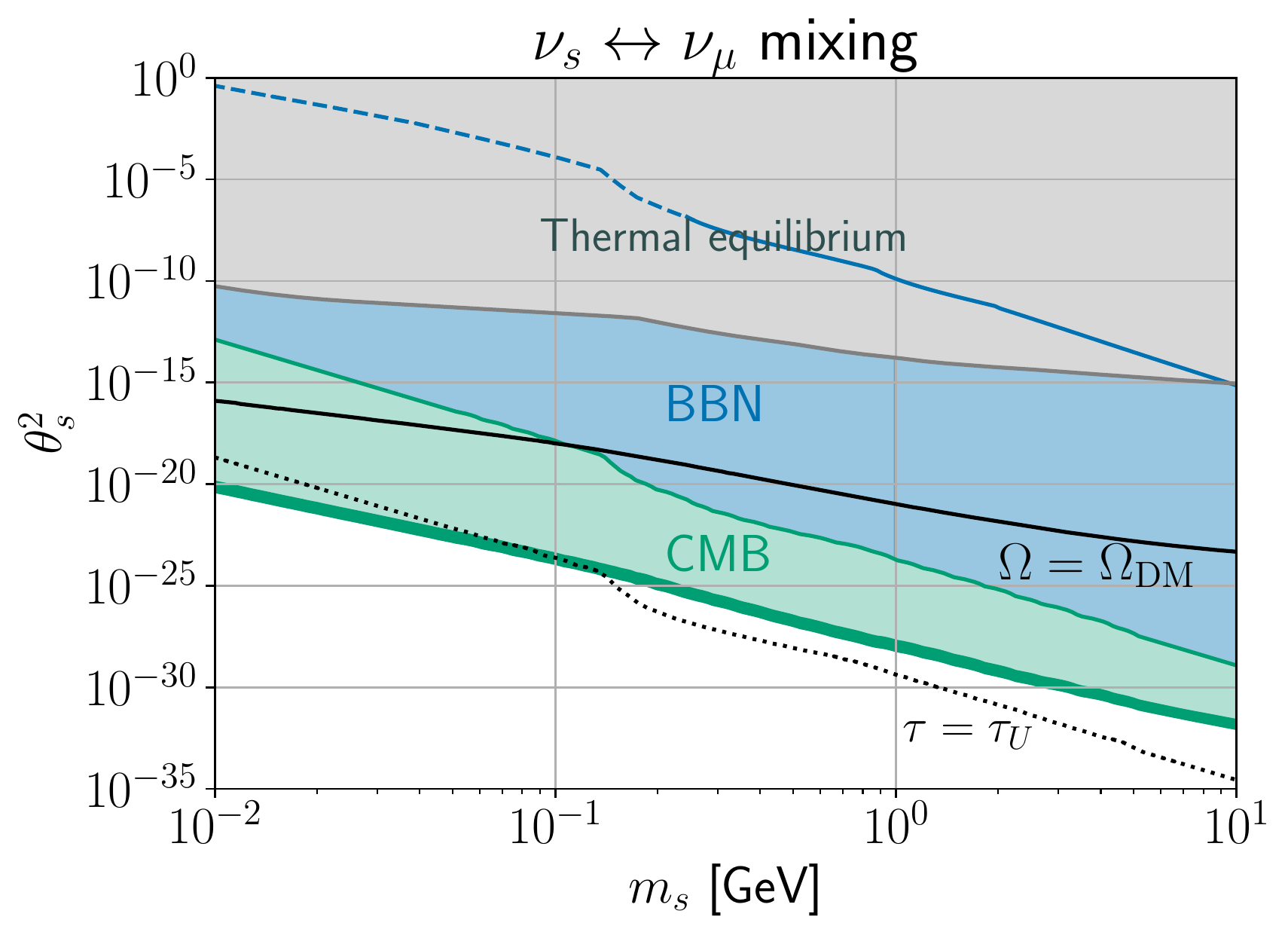}
 \includegraphics[width=0.45\linewidth]{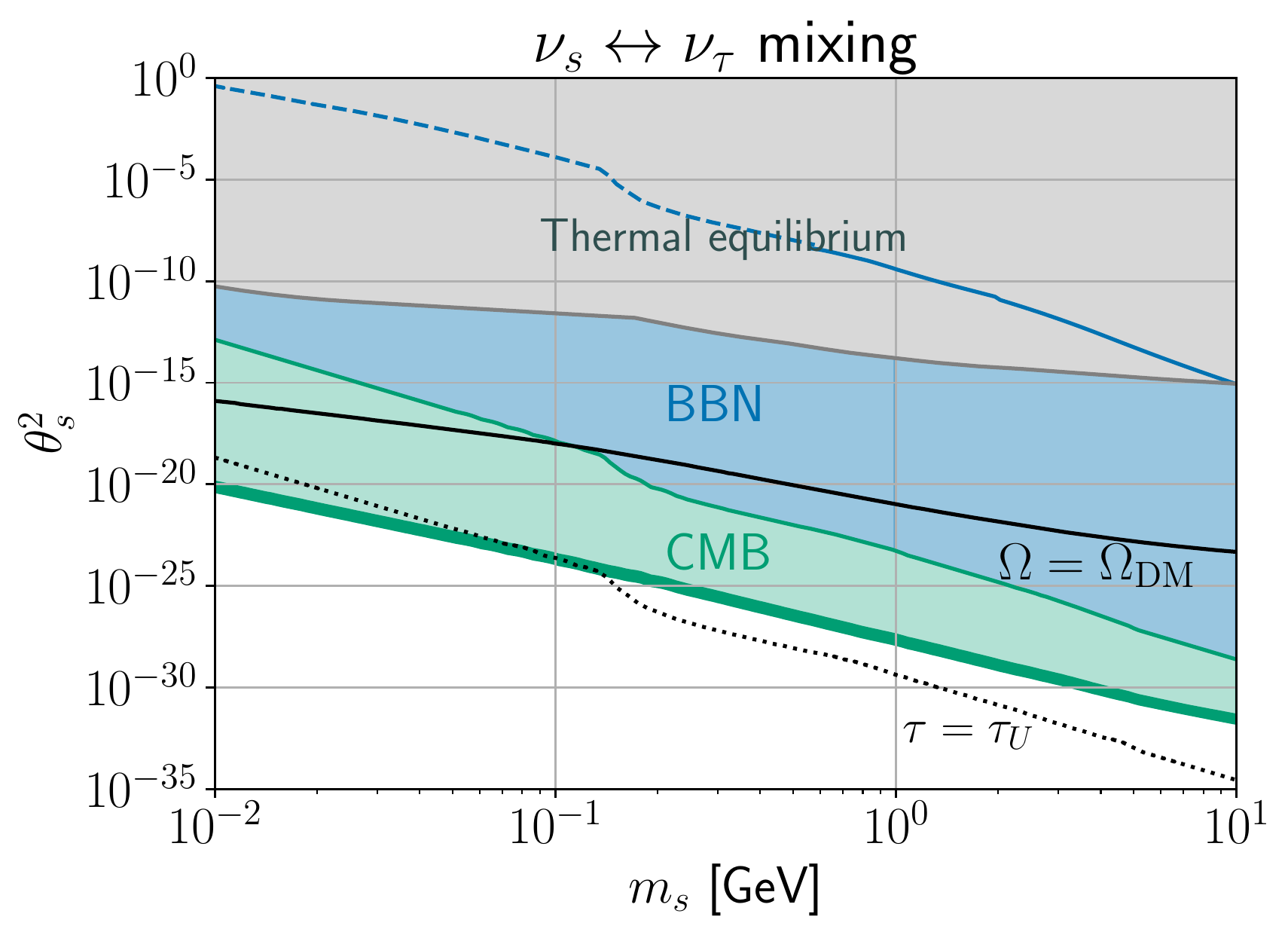}
 \caption{Same as Fig.~\ref{fig:bbn_e} but for sterile neutrinos mixing with muons (left) and tau leptons (right).}
 \label{fig:bbn_mu_tau}
\end{center} 
\end{figure*}

For sufficiently large mixings, the sterile neutrinos become so short-lived that their decays have no impact on BBN.
For sufficiently large values of $m_s$ such that hadronic decays are possible, the upper reach in $\theta_s$ of the constraints can be calculated following Ref.\ \cite{Boyarsky:2020dzc,Bondarenko:2021cpc}.
Our results differ from those in~\cite{Boyarsky:2020dzc,Bondarenko:2021cpc} in that we are interested in sterile neutrinos produced through oscillations rather than thermal processes.
The abundance relevant for our case can be computed using Eq.~\eqref{Ieq}, which leads to the upper boundary of the
BBN exclusion region, shown as the solid blue line in Fig.~\ref{fig:bbn_e} for $e$-mixed $\nu_s$, and Fig.~\ref{fig:bbn_mu_tau} for $\mu$- and $\tau$-mixed $\nu_s$.

At low masses where hadronic decays of sterile neutrinos are forbidden, the predictions of BBN can still be affected by the electromagnetic energy injection and the modification of the expansion history.
In this regime, a full numerical simulation of BBN including sterile neutrinos is needed in order to precisely determine the reach of the exclusion region.
Although this is beyond the scope of the present work, we make a reasonable estimate by adapting the results obtained for thermally produced neutrinos~\cite{Dolgov:2000jw,Dolgov:2000pj,Dolgov:2003sg,Ruchayskiy:2012si,Boyarsky:2020dzc,Gelmini:2020ekg,Sabti:2020yrt} to the nonthermal production mechanism of interest here.
A simple rescaling of the results of~\cite{Sabti:2020yrt} to account for the different $\nu_s$ yields is found to give very good agreement in the region where hadronic decays dominate.
We thereby extend the limits to lower masses where the dominant decays are electromagnetic, resulting in the dashed blue line in Fig.~\ref{fig:bbn_e} for $e$-mixed $\nu_s$ (see Fig.~\ref{fig:bbn_mu_tau} for $\mu$- and $\tau$-mixed $\nu_s$).

Our derivation of the sterile neutrino yield assumes that $\nu_s$ is produced only through oscillations, and not by thermal processes.
This requires that the rate of scatterings such as $\nu_e e \leftrightarrow \nu_s e$ never exceeds the Hubble rate $H$.
The $\nu_s$ thermal production rate, taking into account medium effects, can be estimated as
\begin{equation}
    \bar{\Gamma}_{\rm prod} \simeq \int \mathop{\diff t} \Gamma_a \sin^2(\theta_m) \mathrm{e}^{-\Gamma_a t},
\end{equation}
in terms of the active neutrino scattering rate and the matter mixing angle.
The time average takes into account the widening of the resonance in the matter mixing angle due to elastic scatterings, as in Eq.~\eqref{eq:averaged_oscillation_rate}.
The region where thermal equilibrium is not reached can be identified by requiring that $\bar{\Gamma}_{\rm prod} < 3H$ for all temperatures and neutrino momenta.
In Fig.~\ref{fig:bbn_e} (and Fig.~\ref{fig:bbn_mu_tau}), the region where thermal equilibrium is attained is shaded in grey: our results do not apply for those masses and mixings. 
The relevance of the blue curve in Fig.~\ref{fig:bbn_e} is that there is no gap between the region constrained by BBN and that where thermal processes become important.

In summary, we conclude that CMB and BBN limits rule out the existence of sterile neutrinos with masses $10\,\mathrm{MeV}\leq m_s \leq 10\,\mathrm{GeV}$ and mixings down to $\theta_s\sim 10^{-16}-10^{-10}$, depending on $m_s$.
For larger mixings, sterile neutrinos are thermally produced in the early universe and also constrained by BBN; see e.g.~\cite{Boyarsky:2020dzc,Sabti:2020yrt}.
Those limits take over in the grey region of Fig.~\ref{fig:bbn_e} (and Fig.~\ref{fig:bbn_mu_tau}).

\medskip
{\bf Dark matter production.}  The fast decays of $\nu_s$ 
prevent it from being dark matter. 
However, it is possible that $\nu_s$ decays predominantly into a lighter sterile neutrino $\nu_{s'}$ of mass $m_{s'}$, which can stable on cosmological time scales.
This can alleviate the cosmological constraints by reducing the branching ratio of $\nu_s$ into visible states, 
and it can give rise to a nonthermal mechanism for producing the new DM candidate.  To find the mixing angle needed to get the
right relic density through this mechanism, one should rescale
the mixing angle given by the heavy black curve in  Fig.\ \ref{fig:bbn_e} by $\theta_s\to \theta_s\sqrt{m_{s}/m_{s'}}$ to compensate for the smaller DM mass.

The required decays could arise in a Majoron model \cite{Chikashige:1980ui} involving several sterile neutrinos and two scalar fields that get $(B-L)$-breaking vacuum expectation values (VEVs), with Yukawa couplings
\be
    \mathcal{L}\supset\sfrac12\nu_{s,i}(f_{ij}\phi_1 + g_{ij}\phi_2)\nu_{s,j}\,.
\ee
When the scalars get VEVs $v_i$, the $\nu_s$ mass matrix $m_{s,ij} = v_1 f_{ij} + v_2 g_{ij}$ is generated, but unlike in single-field models, the couplings of the Majoron $J$ need not be diagonal in the mass eigenbasis, leading to $\nu_{s}\to \nu_{s'} J$ decays.  If $\nu_{s'}$ mixes sufficiently weakly to $\nu_a$, it can be a viable DM candidate.

As a proof of concept, consider the case of 
uncoupled potentials for the two scalars,
$V = \sum_i\lambda(|\phi_i|^2-w_i^2)^2$, leading to two separate Majorons, $\phi_i \to w_i e^{i J_i/w}$.  Further assume for simplicity that the Lagrangian states $\nu_{s,1}$ and $\nu_{s,2}$ couple to the single active neutrino species  $\nu_e$, with mass term
and Majoron couplings
\bea
& \sfrac12& \left(\begin{array}{c} \bar\nu_e\\\bar\nu_{s,1}\\\bar\nu_{s,2}\end{array}\right)^{\sss \!\!\!T}
\!\!\!
    \left(\begin{array}{ccc} 0 & 0 & \delta m\\
        0 & f_{1}(w_1+iJ_1) & g(w_2+iJ_2) \\
        \delta m & g(w_2+iJ_2) & f_{2}(w_1+i J_1) \end{array}\right)
        \!\!\!\left(\begin{array}{c} \nu_e\\\nu_{s,1}\\\nu_{s,2}\end{array}\right)\,\nn\\
        &\!\!\!\!\!\to& 
   \sfrac12 \left(\begin{array}{c} \bar\nu_e\\\bar\nu_{s'}\\\bar\nu_{s}\end{array}\right)^{\sss \!\!\!T}
   \!\!\!
    \left(\begin{array}{ccc} 0 & \psi\,\delta m & \delta m\\
        \!\!\!\psi\,\delta m & f_{1}(w_1+iJ_1) & i g\tilde J \\
        m & i g\tilde J & f_{2}(w_1+i J_1)\!\! \end{array}\right)\!\!\!
        \left(\begin{array}{c} \nu_e\\\nu_{s'}\\\nu_{s}\end{array}\right)\,,
        \nn
\eea
with $\gamma_5$ understood to accompany the interaction terms. In the second line we have diagonalized the mass submatrix of the sterile neutrino sector, and denoted $\psi = g w_2/[w_1(f_1-f_2)]\ll 1$ the mixing angle between the sterile neutrinos and $\tilde J = J_2-(w_2/w_1)J_1$.

For concreteness, suppose that $m_{s'} \simeq f_1 w_1 = 0.1\,$GeV and $m_s \simeq f_2 w_1 = 1\,$GeV.  The mixing angles with $\nu_a$ are $\theta_s \simeq \delta m/m_s$ and $\theta_{s'} \simeq \psi\,\delta m/m_s' \simeq \psi (f_2/f_1)\theta_s$.
To get the desired relic density for $\nu_{s'}$, Fig.\ \ref{fig:bbn} requires $\theta_s \simeq  10^{-10}$.  To circumvent the CMB bound 
on $\nu_s$ decays, the rate for $\nu_s\to J_2 + \nu_{s'}$ must be sufficiently greater than the weak rate (\ref{Gweak}), 
\be
    {g^2 m_s\over 16\pi} \gtrsim 10^{8}\,\Gamma_{\rm weak}\,,
\ee
leading to $g \gtrsim 10^{-11}$ for $m_s=1$\,GeV.  On the other hand,
the CMB constraint on $\nu_{s'}$ requires $\theta_{s'}  \lesssim 10^{-16}$ from Fig.\ \ref{fig:bbn} (to obtain $\tau_{s'}\gtrsim 10^{24}\,$s),
hence $\psi \lesssim 10^{-7}$.  In addition, $\nu_{s'}\to J_i \nu_a$ is constrained by limits on 
 monochromatic neutrino lines
\cite{PalomaresRuiz:2007ry,Coy:2020wxp},
\bea
    \Gamma(\nu_s'\to \nu_a+J) &=& {f_1^2\, \theta_{s'}^2\, m_{s'}\over 16\pi} < {1\over 3\times 10^{23}\,{\rm s}}\nn\\
        &\Rightarrow& f_1\theta_{s'} \lesssim 10^{-23}\,.
\eea
All constraints can be satisfied by choosing, for example,
\bea
    f_1 &=& 10^{-7},\ f_2 = 10^{-6},\ g = 10^{-10},\nn\\
        w_1 &=& 10^3\, w_2 = 10^{6}\,{\rm GeV},\ \delta m = 0.1\,{\rm eV}\,.
\eea
After fully diagonalizing the mass matrix, the light neutrino mass is predicted to be
$m_1 = \theta_s^2 m_s \sim 10^{-11}\,$eV.   

\bigskip
{\bf Conclusions.}  Sterile neutrinos at the GeV scale have been the subject of intense study for explaining
baryogenesis via leptogenesis \cite{Asaka:2005pn,Ghiglieri:2017gjz,Ghiglieri:2017csp,Ghiglieri:2018wbs,Klaric:2020lov,Bondarenko:2021cpc}.
In this work we have shown that resonant active-sterile neutrino oscillations can efficiently produce GeV-scale $\nu_s$ in the early universe, without requiring any additional new physics like the presence of lepton asymmetries.

The decays of resonantly produced sterile neutrinos can have a significant impact on BBN and the CMB.
This has allowed us to exclude mixing angles up to nine orders of magnitude smaller than had been considered in previous analyses based on thermal production processes.
Thus, GeV-scale leptogenesis remains viable only for large active-sterile mixings for which the $\nu_s$ lifetime becomes short enough to make it harmless for BBN.

We have also proposed a novel mechanism by which resonant production of GeV-scale $\nu_s$ could lead to heavier-than-MeV sterile neutrino dark matter through decays of the originally-produced $\nu_s$.

The nonequilibrium density matrix formalism \cite{Stodolsky:1986dx} provides a
more rigorous framework for treating the $\nu_s$ production mechanism studied here.
While we have confirmed that our simpler and less numerically demanding approach agrees with this method in the nonresonant regime, it would be desirable to study the resonant conversions within this more sophisticated approach.
We leave this task for future work.

\medskip
{\bf Acknowledgment.} 
We thank K.\ Kainulainen and M.\ Laine for helpful discussions.
This work was supported by NSERC (Natural Sciences
and Engineering Research Council, Canada).
G.A. is supported by the McGill Space Institute through a McGill Trottier Chair Astrophysics Postdoctoral Fellowship.
\medskip

{\bf Appendix A: thermal neutrino self-energies.}
The thermal contribution to the $\nu_a$ self-energy can be found in~\cite{Quimbay:1995jn}. 
In the relativistic limit, the dispersion relation can be parametrized\footnote{We only consider modifications in the $b_L$ functions but neglect ones in $a_L$.} as $\omega = k + V_a$, with
\bea
    V_a&=& -\left\{\begin{array}{ll} v_\Z {B}(0,m_\Z) + v_\W{B}(m_\ell,m_W),& \mathrm{broken\ phase}\\
        (v_\Z + v_\W){B}(0,0),& \mathrm{sym.\ phase}
        \end{array} \right.\nn\\
        \label{bleq}
\eea 
where $v_\Z = \pi\alpha_\W / c_\W^2$ and $v_\W = \pi\alpha_\W (2 + m_\ell^2/m_\W^2)$. 
The function $B$ depends on the neutrino energy $\omega$ and momentum $p$, as well as $T$, $m_{\W,\Z}$ and $m_{\ell}$.
Treating the thermal contribution as a perturbation, we can set $\omega\simeq p$, the unperturbed on-shell relation,\footnote{It might be questioned whether $\omega = \sqrt{m_s^2+p^2}$ is the more appropriate choice for production of on-shell $\nu_s$ \cite{Asaka:2006nq}.  We have checked that the ambiguity has no impact on our results.} which simplifies the form of ${B}(m_f,m_A)$ to
\bea \label{eq:B0mZ}
    {B}(m_f,m_\A) &=& {1\over p^2}\int {dk\over 8\pi^2}\Bigg[
        \left({\Delta\over 2} \frac{k}{\epsilon_\A} L_2^+(k) - 4{pk^2\over \epsilon_\A}\right)n_b(\epsilon_\A) \nn\\
        &+&\left({\Delta\over 2}\frac{k}{\epsilon_f} L_1^+(k) - \frac{4pk^2}{\epsilon_f}\right)n_f(\epsilon_f)\Bigg],
\eea
where $\Delta = m_\A^2-m_f^2$, $n_f$ and $n_b$ are the Fermi-Dirac or Bose-Einstein distribution functions, $\epsilon_i = \sqrt{k^2 + m_i^2}$, and
\bea
    L_2^+(k) &=& \ln\left(\left|[\Delta + 2p(k+\epsilon_\A)][\Delta + 2p(k-\epsilon_\A)]\over
    [\Delta - 2p(k-\epsilon_\A)][\Delta - 2p(k+\epsilon_\A)]\right|\right)\nn\\
    L_1^+(k) &=& \ln\left(\left|[\Delta + 2p(k+\epsilon_f)][\Delta + 2p(k-\epsilon_f)]\over
    [\Delta - 2p(k-\epsilon_f)][\Delta - 2p(k+\epsilon_f)]\right|\right).\quad\ 
\eea

After the integration, $B$ becomes a function of $p$ only, as has been assumed in the
text. 
A smooth crossover at the electroweak phase transition can be achieved by setting $m^2_{\Z,\W}(T) \simeq
m^2_{\Z,\W} (1 - T^2/T_{\rm EW}^2)$.
At high temperatures, $V_a$ can be further simplified by taking $m_{\ell} \simeq 0$ in Eq.\ (\ref{bleq}).  At temperatures $T\ll T_{\rm EW}$, the integral~\eqref{eq:B0mZ} can be performed analytically, leading to Eq.\ (\ref{eq:vweq}).

The resulting matter potential is shown in Fig.\ \ref{fig:Va} as a function of the temperature, for some fixed values of $p/T$.
The change of sign of $V_a$, as well as the limiting behaviours in Eq.\ \eqref{eq:vweq}, can be appreciated there.

\medskip
{\bf Appendix B: limits for $\mu$- and $\tau$-mixed $\nu_s$.}\\
Fig.~\ref{fig:bbn_mu_tau} provides the analogous results to Fig.~\ref{fig:bbn_e} for the case of $\nu_s$ mixing with $\nu_\mu$ or $\nu_\tau$, respectively.

\bibliography{ref}
\bibliographystyle{utphys}
\end{document}